\magnification=\magstep1

\input epsf
\def\Narrower{\par\narrower\noindent}

\baselineskip=20 truept
\def\cl{\centerline} 
\def\ni{\noindent}
\def\hb{\hfil\break}
\def\bn{\bigskip\noindent}
\def\mn{\medskip\noindent}

\def\ie{{\it i.e.}}
\def\pa{\partial}

\def\tr{\mathop{\rm tr}\nolimits}

\def\half{{\textstyle{1\over2}}}

\def\UH{\widehat U}
\def\ii{{\rm i}}  
\def\ee{{\rm e}}  
\def\vp{\varphi}
\def\ve{\varepsilon}
\font\scap=cmcsc10
\font\title=cmbx10 scaled\magstep1

\def\a{\alpha}
\def\b{\beta}
\def\g{\gamma}

\def\s{\sigma}
\def\z{\zeta}

\def\MT{\widetilde M}
\def\vp{\varphi}
\def\th{\theta}
\def\vpt{\tilde\varphi}
\def\tht{\tilde\theta}
\def\cosec{\mathop{\rm cosec}\nolimits}
\def\sech{\mathop{\rm sech}\nolimits}
\def\tanh{\mathop{\rm tanh}\nolimits}

\def\wb{\bar w}
\def\ub{\bar u}
\centerline{\title Integrable Yang-Mills-Higgs Equations}
\centerline{\title in 3-Dimensional De Sitter Space-Time.}

\vskip 1truein
\centerline{\scap V.\ Kotecha and R.\ S.\ Ward}

\bn\centerline{\it Department of Mathematical Sciences, University of Durham,}
\centerline{\it Durham DH1 3LE, UK.}

\vskip 2truein
\ni{\bf Abstract.} This paper describes an integrable Yang-Mills-Higgs
system on
\hb
(2+1)-dimensional de Sitter space-time.  It is the curved-space-time
analogue of the Bogomolnyi equations for monopoles on ${\bf R^3}$.  A number
of solutions, of various types, are constructed.

\vfil\eject


\bn1. {\scap Introduction.}
\mn The background to this paper is the question of the existence of
integrable nonlinear partial differential equations (and more specifically
of soliton equations) in curved space-times.  For a given (fixed) space-time
$M$, are there integrable systems which live on $M$ (\ie\ are covariantly
coupled to its geometry)?  In general, this places severe restrictions
both on $M$ and on the equations that are coupled to it.  In this paper,
we concentrate on one example, namely an integrable Yang-Mills-Higgs
system on (2+1)-dimensional de Sitter space-time.

In effect, this generalizes examples which have long been known.
Consider the chiral equation $g^{\mu\nu}\pa_\mu(U^{-1}\pa_\nu U) = 0$,
where $U(x^\mu)$ takes values in a Lie group, and where $g_{\mu\nu}$ is
the metric of $M$.  This system is integrable if $M$ is (1+1)-dimensional
(this being related to conformal invariance).
In higher-dimensional flat space-times, the chiral equation is not
integrable$^1$; and this is probably also the case for curved space-times
of dimension greater than two.  But if one modifies the equation by
adding a torsion term, then integrability is possible$^{1,2}$; in particular,
there is an integrable (modified) chiral equation in flat 3-dimensional
space-time ${\bf R}^{2+1}$.  The system is equivalent to one involving a gauge
field (Yang-Mills field) coupled to a Higgs field, and may be seen to arise
from the self-dual Yang-Mills equations in ${\bf R}^{2+2}$, by dimensional
reduction.  The soliton solutions can be understood in terms of algebraic
geometry, and the soliton dynamics is (in general non-trivial$^{3-10}$.

Other ways of reducing the self-dual Yang-Mills equations in ${\bf R}^{2+2}$
can lead to integrable Yang-Mills-Higgs systems in curved (2+1)-dimensional
space-times.  These are the Lorentzian analogue of hyperbolic monopoles,
which live on (positive-definite) hyperbolic 3-space.  The space-time
has to have constant curvature; and so there are two Lorentzian
possibilities, namely anti-de Sitter and de Sitter space-time.
Some preliminary results on the anti-de Sitter case have appeared
previously$^{11}$; the present paper deals with the de Sitter case.
In particular,
we construct various explicit solutions.  One new feature
that appears here is  associated with the non-trivial topology of de Sitter
space.

\bn2. {\scap (2+1)-Dimensional De Sitter Space-Time.}

\mn (2+1)-dimensional de Sitter space-time $M$ is the manifold
${\bf R}\times S^2$ equipped with the metric
$$
ds^2 = g_{\mu\nu}\,dx^\mu\,dx^\nu
       = \cosh^2 T (d\th^2 + \sin^2\th\,d\vp^2) - dT^2. \eqno(1)
$$
Here $T\in{\bf R}$ is a time coordinate, and $(\th,\vp)$ are polar coordinates
on the spatial sphere.  It is a space of constant curvature, with scalar
curvature $R=6$ (conventions are those of ref 12).

There is a relation between this space-time $M$ and flat (2+2)-dimensional
space ${\bf R}^{2+2}$, and we shall use this to obtain equations on $M$ from
equations on ${\bf R}^{2+2}$.  The relation is as follows.  Let $u$ and $w$ be
complex coordinates on ${\bf R}^{2+2}$, so that its metric is
$du\,d\ub - dw\,d\wb$.
First, define new coordinates $(\th,\vp,\tht,\vpt)$ by
$$
u = {(\sin\th)\ee^{-\ii\vp}\over(\cos\th+\cos\tht)}, \quad
w = {(\sin\tht)\ee^{\ii\vpt}\over(\cos\th+\cos\tht)}. \eqno(2)
$$
Then
$$
du\,d\ub - dw\,d\wb = 2(\cos\th+\cos\tht)^{-1}\,ds^2_{\MT}, \eqno(3)
$$
where
$$
ds^2_{\MT} = (d\th^2 + \sin^2\th\,d\vp^2) - (d\tht^2 + \sin^2\tht\,d\vpt^2). 
      \eqno(4)
$$
In other words, $R^{2+2}$ is conformal to part of the product
$\MT = S^2\times S^2$;
we interpret $(\th,\vp)$ as polar coordinates on the first sphere of $\MT$,
and $(\tht,\vpt)$ on the second. The 4-space $\MT$ is conformally flat and
has vanishing scalar curvature;
it is a double cover of a conformal compactification$^{13,14}$ of
${\bf R}^{2+2}$.

The next step is to reduce to 2+1 dimensions: this is done by
factoring out by the Killing vector $\pa/\pa\vpt$, \ie\ by a rotation
of the second sphere. First we remove $\tht = 0$ and $\tht = \pi$, which
are fixed points of the rotation.  On the complement of these fixed points,
we can write
$$
ds^2_{\MT} = \sin^2\tht\bigl[\cosec^2\tht(d\th^2 + \sin^2\th\,d\vp^2 - d\tht^2)
        - d\vpt^2 \bigr]. \eqno(5)
$$
So $\MT$ (minus the fixed points) is conformal to the product of $S^1$
and a space with topology  ${\bf R}\times S^2$ and metric
$$
ds^2 = \cosec^2\tht(d\th^2 + \sin^2\th\,d\vp^2 - d\tht^2). \eqno(6)
$$
This is exactly (2+1)-dimensional de Sitter space-time (1), where the
coordinates $T$ and $\tht$ are related by
$\tanh T = -\cos\tht$.


\bn3. {\scap Integrable Equations on $M$.}
\mn In view of the conformal relation between $\MT$ and $M$, we may
obtain integrable equations on $M$ by reducing conformally-invariant
integrable equations on $\MT$ (or ${\bf R}^{2+2}$).
The simplest conformally-invariant equation on $\MT$ is the
conformally-invariant wave equation.  Bearing in mind the absence of scalar
curvature, this has the form
$$
  \Delta \chi - \widetilde\Delta \chi = 0, \eqno(7)
$$
where $\Delta$ and $\widetilde\Delta$ are the Laplacians on the two spheres.
The $\vpt$-independent solutions of (7) correspond to solutions of the
conformally-invariant wave equation on $M$, namely
$$
  g^{\mu\nu} \nabla_\mu \nabla_\nu \Psi - \Psi = 0,  \eqno(8)
$$
where $\chi$ and $\Psi$ are related by the relevant conformal factor:
$\Psi = (\sech T) \chi$.
Solutions can be obtained (in terms of Legendre polynomials and
spherical harmonics) by separating variables or by twistor methods$^{13}$.
For example, the simplest case $\chi = 1$ (constant) gives $\Psi = \sech T$,
\ie\ a solution of (8) which is spatially constant.
Using ${\it l} = 1$ spherical
harmonics yields the examples $\Psi = \sech T\,\tanh T\, \cos\th$,
$\Psi = \sech T\,\tanh T\, \sin\th\, \cos\vp$ etc.

Another example, and the one which we concentrate on in this article, is
that of the self-dual Yang-Mills equations (these are integrable on any
conformally-flat 4-space, and so in particular on $\MT$).
When we reduce to the (2+1)-dimensional space-time $M$, the self-dual
Yang-Mills field becomes a Yang-Mills-Higgs system
$(\Phi, A_\mu)$ satisfying the Bogomolny-type equations
$$
   D_\a\,\Phi = \half \eta_{\a\b\g} F^{\b\g}.  \eqno(9)
$$
The Higgs field $\Phi$ (taking values in the Lie algebra $G$ of the gauge
group) is identified with the $\vpt$-component $A_{\vpt}$
of the gauge potential, with the remaining three components $A_\mu$ becoming
a gauge potential on $M$.  As usual, $D_\a$ denotes the covariant derivative
$D_\a \Phi = \pa_\a \Phi + [A_\a,\Phi]$, $F_{\mu\nu}$ is the gauge field
$[D_\mu,D_\nu]$, and
$\eta_{\a\b\g} = [-\det(g_{\mu\nu})]^{1/2}\,\ve_{\a\b\g}$ is the
volume 3-form on $M$.  In terms of the polar coordinates $(\th,\tht,\vp)$,
eqn (9) is
$$\eqalign{
  D_{\tht} \, \Phi &= (\sin\tht/\sin\th)\, F_{\th\vp}, \cr
  D_{\th} \, \Phi &= (\sin\tht/\sin\th)\, F_{\tht\vp}, \cr
  D_{\vp} \, \Phi &= (\sin\tht\sin\th)\, F_{\th\tht}. \cr
}\eqno(10)$$

So (9), or equivalently (10), form a set of covariant integrable partial
differential equations on $M$.  They are linear if the gauge algebra
$G$ is abelian, but otherwise are nonlinear.
In the remaining sections, we shall construct and examine some solutions
of (10), for gauge algebras $u(1)$ and $su(2)$.


\bn4. {\scap A U(1) example.}
\mn For gauge algebra $G = u(1)$, the equations (10) reduce to
$$
\pa_\a \Phi = \half\eta_{\a\b\g} F^{\b\g}, \eqno(11)
$$
where $F_{\mu\nu} = \pa_{\mu} A_{\nu} - \pa_{\nu} A_{\mu}$.
Note that from (11) it follows immediately that
$g^{\mu\nu} \nabla_\mu \nabla_\nu \Phi = 0$;
so this case is related to, but different from, that of the wave 
equation (8) discussed previously.  Since space is a sphere
$S^2$, there can be non-trivial topology: U(1) gauge fields over $S^2$ are
classified topologically by the integer
$$
k = {-\ii\over2\pi} \int_{\Sigma}F_{\mu\nu} \, dx^\mu \wedge dx^\nu,\eqno(12)
$$
where $\Sigma$ is a space section (spacelike surface with topology $S^2$).

An example of a topologically non-trivial solution of (11) is
$$
  \Phi=\half\ii k(\cos\tht-1), \quad A_{\vp}=\half\ii k(\cos\th-1),
    \quad A_{\th}=0= A_{\tht}.
$$
For smoothness, we need $A_{\vp}=0$ at $\th=0,\pi$; so the above
gauge potential has a singularity at $\th=\pi$.  This is the familiar
`Dirac-string' singularity, and is removable: the gauge-transformed
potential
$$A_{\vp} + \exp(-\ii k\vp)\pa_{\vp}\exp(\ii k\vp)
     = \half\ii k(\cos\th+1)
$$
is smooth near $\th=\pi$.  In other words, this Maxwell-Higgs system is smooth
throughout $M$.  The apparent singularities are a consequence of the fact
that the gauge field is topologically non-trivial: its magnetic charge
equals $k$.  Furthermore, it is spatially-homogeneous: note in particular that
$\Phi$ depends only on the time coordinate $\tht$, and that the gauge 2-form
(the integrand
of (12)) is a (time-dependent) multiple of the spatial area element
$\sin\th\,d\th\wedge d\vp$.


\bn5. {\scap Spatially-Homogeneous SU(2) Solutions.}
\mn Spatially-homogeneous SU(2) fields may be characterized as follows.
Temporarily, think of the
spatial 2-sphere as the unit sphere in ${\bf R}^3$, with coordinates
$x^j = (x^1,x^2,x^3)$.  Take the Higgs field and gauge potential to have
the form
$$\eqalign{
  \Phi &= \ii g(\tht) x^j \s_j, \cr
  A_j &= \ii f(\tht) \ve_{jkl}x^l\s^k, \cr
  A_{\tht} &= 0, \quad{\rm (a\ gauge\ choice)} \cr
  }\eqno(13)
$$
where $\s_j$ are the the Pauli matrices, and $f$ and $g$ are two scalar
functions of $\tht$ only.  This implements SO(3) symmetry: recall, for
example, that the
spherically-symmetric 1-monopole in ${\bf R}^3$ has the `hedgehog' form (13).
The components $A_{\th}$ and $A_{\vp}$ are
obtained from $A_j$ in the obvious way, by transforming to polar coordinates.
Although $\Phi$ and $A_\mu$ depend on the spatial variables $(\th,\vp)$,
the effect of a spatial rotation is to make a gauge transformation; and
gauge-invariant quantities such as $-\tr\Phi^2 = 2 g^2$ depend only on $\tht$.

Substituting (13) into (10) gives the pair of ordinary differential equations
$$\eqalign{
  g' &= 2 f(1-f)\sin\tht, \cr
  f' &= g(2f-1)/\sin\tht. \cr
  }\eqno(14)
$$
Eliminating $g$ from these leaves an equation for $f$ which, after the
transformation
$$
f(\tht) = \half(\ee^{2T}+1) P(T) + \half,\quad \tanh T = -\cos\tht,
$$
is
$$
  P'' = (P')^2/P - 4 \ee^{2T} P^3,  \eqno(15)
$$
where $P' = dP/dT$.
This is the third Painlev\'e equation $P_{III}$.  In terms of the
variable $t = \ee^T = \tan(\tht/2) \in(0,\infty)$, it takes the more usual form
$$
 \ddot P = (\dot P)^2/P - \dot P/t - 4 P^3, \eqno(16)
$$
where $\dot P = dP/dt$.
Solutions of (15) or (16) therefore determine spatially-homogeneous SU(2)
solutions of the Yang-Mills-Higgs-Bogomolny equations (10).


\bn6. {\scap The Twistor Correspondence.}
\mn One can in principle construct all solutions of the self-dual Yang-Mills
equations, and hence of (10), by using the twistor correspondence$^{15,14}$.
The details of the construction are well-known,
and here we simply give some brief details in order to establish notation
and conventions.

Twistor space is the complex projective space ${\bf CP^3}$, with homogeneous
coordinates $Z^\a = (Z^0,Z^1,Z^2,Z^3)$.  (Strictly speaking, the twistor
space of $\MT$ is a non-Hausdorff space$^{14}$ obtained by glueing together
two copies of ${\bf CP^3}$; but for simplicity we shall avoid going
into the details of this.)
The correspondence between  ${\bf CP^3}$ and $\MT$ is expressed by the
relations
$$
Z^0 = uZ^2 + wZ^3, \quad Z^1 = \wb Z^2 + \ub Z^3. \eqno(17)
$$
Here $u$ and $w$ are the complex coordinates defined by (2) (recall that
they only cover `half' of $\MT$ --- it is for this reason that the true twistor
space is a non-Hausdorff `doubling' of ${\bf CP^3}$).

A matrix-valued twistor function $F(Z^\a)$ is said to be {\it real} if
$F^{\dagger} = F$, where $F^{\dagger}(Z^\a) 
 = F(\overline{Z^1},\overline{Z^0},\overline{Z^3},\overline{Z^2})^*$,
and $^*$ denotes complex conjugate transpose.  There is a correspondence
between certain holomorphic vector bundles over twistor space, and solutions
of the self-dual Yang-Mills equations on $\MT$; in particular,
if $F(Z^\a)$ is a real `patching matrix' for a vector bundle of rank $n$,
then `splitting' $F$ yields a self-dual U($n$) gauge field.
In addition to being real, the matrix function $F(Z^\a)$ has to be
homogeneous of degree zero in $Z^\a$; and in order to have $\vpt$-invariance,
we require $F$ to be annihilated by the vector field
$$
  V = Z^3{\pa\over\pa Z^3} - Z^2{\pa\over\pa Z^2}
    +  Z^1{\pa\over\pa Z^1} - Z^0{\pa\over\pa Z^0}. \eqno(18)
$$
For example, all three requirements (reality, homogeneity and $V$-invariance)
are met by the (scalar) function $Q = (Z^0 Z^1 + Z^2 Z^3)/(Z^2 Z^3)$.
Indeed, the line bundle defined by the patching matrix
$F = Q^k$, where $k$ is an integer, yields the U(1) solution of Section 4.


\bn7. {\scap An SU(2) Example.}
\mn In order to obtain SU(2) solutions by this construction, we look for
examples of
$2\times2$ twistor matrices $F(Z^\a)$ which are upper-triangular, and
which are equivalent to `real' matrices.  Given an upper-triangular $F$,
one can obtain explicit expressions
for $\Phi$ and $A_{\mu}$ (see, for example,  section 8.2 of ref 15).
The analogue of the 'tHooft ansatz, and its generalizations (corresponding
to example 8.2.3 of ref 15) does not work --- it produces only SU(1,1)
fields.  But the analogue (changed-signature version) of example 8.2.4 of
ref 15 does work, and produces SU(2) solutions in our case.  Some brief
details are as follows.

Write $\z = Z^3/Z^2$, and think of $F(Z^\a)$ as defining a vector bundle
by the patching relation $\hat\psi = F \psi$, where $\psi$ and $\hat\psi$
are (2-vector) fibre-coordinates over $U=\{|\z|\leq 1\}$ and
$\UH=\{|\z|\geq 1\}$ respectively.  Take $F(Z^\a)$ to have the form
$$
F(Z^\a) = \pmatrix{\z^k\ee^f & 2Q^{-1}\cosh f \cr 0 & \z^{-k}\ee^{-f} \cr},
          \eqno(19)
$$
where $k$ is a positive integer, $f(Z^\a)$ is real, and $Q = P/(Z^2 Z^3)^k$
with $P(Z^\a)$ being a real polynomial (homogeneous of degree $2k$).
Then, because
$$
R(Z^\a) = \pmatrix{0 & -1 \cr 1 & \z^k Q \cr}
$$
is holomorphic on $U$ and $FR$ is real, it follows that the construction
will yield a real (\ie\ SU(2)-valued) solution.

As an example of this construction, take $P = (Z^0 Z^1 + Z^2 Z^3)$ and
$k=1$ (or $k=-1$, which leads to the same solution).
The simplest choice for $f$, namely $f=0$, gives nothing new: the field
is then effectively abelian, and is an embedding into SU(2) of the
U(1) solution described in section 4.  To get something genuinely non-abelian,
we may take
$f = \log Q$, where $Q = (Z^0 Z^1 + Z^2 Z^3)/(Z^2 Z^3)$, so that
$$
F(Z^\a) = \pmatrix{\z Q & 1+Q^{-2} \cr 0 & (\z Q)^{-1} \cr}. \eqno(20)
$$
The procedure$^{15}$ referred to above then
yields explicit (although rather complicated) expressions
for $\Phi$ and $A_{\mu}$, as rational functions of $\cos\th$, $\cos\tht$
and $\exp(\ii\vp)$.  The dependence on $\vp$ can be compensated by a gauge
transformation, so in effect the solution depends only on $\th$ and $\tht$:
it is an SO(2)-invariant solution of the Yang-Mills-Higgs equations (9) on $M$.

The functions are somewhat simpler when expressed
in terms of the variables $X=\cos^2(\th/2)$ and $Y=\cos^2(\tht/2)$;
for example, $-\tr\Phi^2 = \half H(X,Y)/(1+X^2Y^2)^2$, where
$$
H(X,Y)= 1 + 16X^{4}Y^{6} - 24X^{4}Y^{5} + 9X^{4}Y^{4}
 + 16X^{2}Y^{4} - 8X^{2}Y^{3} - 6X^{2}Y^{2} - 16XY^{4} + 16XY^{3}.
$$
Figure 1 contains plots of four gauge-invariant quantities, namely
$$\eqalign{
  K &:= -\tr\Phi^2, \cr
  L &:= -\sin^2\tht\tr(D_{\tht}\Phi)^2, \cr
  M &:= -\sin^2\tht\tr\bigl[(D_{\th}\Phi)^2
               + (D_{\vp}\Phi)^2/\sin^2\th  \bigr], \cr
  N &:= L-M = g^{\mu\nu}\tr\bigl[ (D_{\mu}\Phi)(D_{\nu}\Phi) \bigr], \cr
}$$
as functions of `spatial latitude' $X$ and `time' $Y$.  A couple of
features that may be noted are:
\item{$\bullet$} in the distant future or past (\ie\ as $Y\to1$ or $Y\to0$),
  the field approaches a `vacuum value' where $-\tr\Phi^2 = \half$
  and  $-\tr(D_\mu\Phi)^2 = 0$;
\item{$\bullet$} at the point $X=0$ on the spatial sphere, we have
  $-\tr\Phi^2 = \half$, $-\tr(D_{\rm time}\Phi)^2 = 0$ and 
  $-\tr(D_{\rm space}\Phi)^2 = 16Y^4(Y-1)^2$.

\midinsert\cl{\epsffile{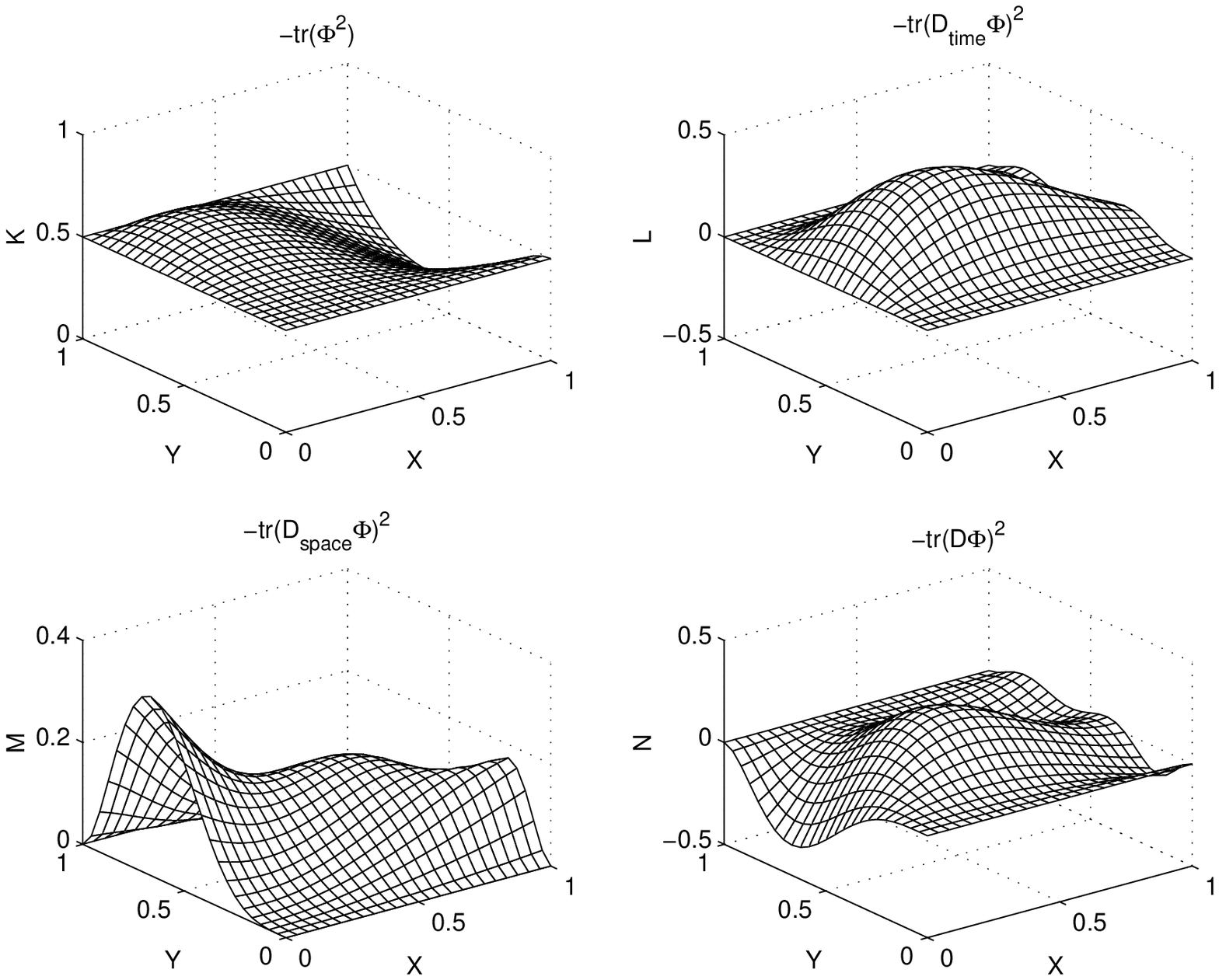}} \Narrower
Fig.~1. The quantities $K=-\tr\Phi^2$, $L=-\sin^2\tht\tr(D_{\tht}\Phi)^2$,
\hb
$M = -\sin^2\tht\tr\bigl[(D_{\th}\Phi)^2 + (D_{\vp}\Phi)^2/\sin^2\th  \bigr]$
and $N =-g^{\mu\nu}\tr\bigl[ (D_{\mu}\Phi)(D_{\nu}\Phi) \bigr]$
as functions of $X=\cos^2(\th/2)$ and $Y=\cos^2(\tht/2)$. \endinsert

\vfill\eject
\bn8. {\scap Concluding Remarks.}
\mn For the corresponding systems in (2+1)-dimensional flat$^2$  and
anti-de Sitter$^{11}$  space-time, there are localized soliton solutions;
and a single soliton travels (as one would expect) along a timelike
geodesic.  More investigation is needed to determine whether the same
is true in the de Sitter case.  The method used to construct solutions
in the former cases does not work so well here; the construction of section 7
is, by contrast, the analogue of one which yields the one-monopole
solution$^{15}$ of the Yang-Mills-Higgs-Bogomolnyi equations on ${\bf R^3}$.
One question, therefore, is whether there is a meaningful correspondence
between between these two systems, \ie\ between the Yang-Mills-Higgs systems
on ${\bf R^3}$ and on (2+1)-dimensional de Sitter space.

In addition to exact solution methods, one may wish to investigate the
equations numerically, as was done in the flat case$^5$.  For this, an
alternative sigma-model or chiral-model formulation is useful; and this
may be of interest in any event.  For example, there exists a gauge in
which $A_{\bar u} = H^{-1}\pa_{\bar u} H$ and $A_w = H^{-1}\pa_w H$ ,
where $H$ takes values in the complexified gauge group (\ie\ SL(2,C)
if $G=su(2)$).  Then the hermitian matrix $K=H H^*$ satisfies
$$
  \pa_u (K^{-1} \pa_{\bar u}K) - \pa_{\bar w} (K^{-1} \pa_w K) = 0. \eqno(21)
$$
And this single matrix equation (21) is equivalent, after transforming
coordinates as in (2) and imposing a suitable dependence on $\vpt$,
to the Yang-Mills-Higgs equations (10).


\bn{\bf References.}
%
\item{$^1$} R.~S.~Ward, Nonlinearity {\bf1}, 671 (1988).
\item{$^2$} R.~S.~Ward, J.~Math.~Phys. {\bf29}, 386 (1988).
\item{$^3$} R.~S.~Ward, Commun.~Math.~Phys. {\bf128}, 319 (1990).
\item{$^4$} P.~M.~Sutcliffe, J.~Math.~Phys. {\bf33}, 2269 (1992).
\item{$^5$} P.~M.~Sutcliffe, Phys.~Rev.~D {\bf47}, 5470 (1993).
\item{$^6$} R.~S.~Ward, Phys.~Lett.~A {\bf208}, 203 (1995).
\item{$^7$} C.~K.~Anand, Commun.~Analysis \&\ Geometry {\bf3}, 371 (1995).
\item{$^8$} T.~Ioannidou, J.~Math.~Phys. {\bf37}, 3422 (1996).
\item{$^9$} R.~S.~Ward, {\it Twistors, geometry and integrable systems.}
        In: {\it The Geometric Universe}, eds S A Huggett et al
        (Oxford University Press, 1998), pp 91-108.
\item{$^{10}$} T.~Ioannidou and W.~J.~Zakrzewski, J.~Math.~Phys.
       {\bf39}, 2693 (1998).
\item{$^{11}$} R.~S.~Ward, Asian J Math {\bf3}, 325 (1999).
\item{$^{12}$} S.~W.~Hawking and G.~F.~R.~Ellis, {\it The Large-Scale Structure
  of Space-Time} (Cambridge University Press, Cambridge, 1973).
\item{$^{13}$} N.~M.~J.~Woodhouse, Proc Roy Soc Lond A {\bf438}, 197 (1992).
\item{$^{14}$} L.~J.~Mason and N.~M.~J.~Woodhouse, {\it Integrability,
   Self-Duality, and Twistor Theory} (Oxford University Press, 1996).
\item{$^{15}$} R.~S.~Ward and R.~O.~Wells, {\it Twistor Geometry and Field
      Theory} (Cambridge University Press, 1990).

\bye